\documentclass[12pt,preprint]{aastex}

\begin{document}

\title {POLYCYCLIC AROMATIC HYDROCARBONS ORBITING HD 233517, AN EVOLVED OXYGEN-RICH RED GIANT\footnote{Based on observations with the NASA
 {\it Spitzer Space Telescope}, which is operated by the California Institute of Technology for NASA}}

\author{M. Jura\footnote{Department of Physics and Astronomy and Center for Astrobiology, University of California,
 Los Angeles CA 90095-1562; jura@astro.ucla.edu}$\,\,$,  C. J. Bohac\footnote{Department of Physics and Astronomy, University of Rochester,
 Rochester NY 14627-0171; cb005k@mail.rochester.edu; bsargent@pas.rochester.edu; joel@pas.rochester.edu; forrest@pas.rochester.edu; dmw@pas.rochester.edu}$\;$, B. Sargent$^{3}\,$ W. J. Forrest$^{3}\,$, J. Green$^{3}$\,,  D. M. Watson$^{3}$, G. C. Sloan\footnote{Department of Astronomy, Cornell University, Ithaca NY 14853-6801; sloan@isc.astro.cornell.edu},
 F. Markwick-Kemper\footnote{Department of Astronomy,
 University of Virginia, Charlottesville VA 22903; fk2n@virginia.edu}$\;\,$, C. H. Chen\footnote{National Optical Astronomy Observatories,
 Tucson AZ 85726-6732; cchen@noao.edu; najita@noao.edu} $^{,}$\,\footnote{{\it Spitzer} Fellow}$\;\;$, J. Najita$^{6}$}

\begin{abstract}
 
We report  spectra obtained with the {\it Spitzer Space Telescope} in the ${\lambda}$ = 5 -- 35 ${\mu}$m range of HD 233517, an
 evolved K2 III giant with circumstellar dust.  For ${\lambda}$ $>$ 13 ${\mu}$m, the flux is a smooth continuum that varies approximately
 as ${\nu}^{-5/3}$. For ${\lambda}$ $<$ 13 ${\mu}$m, although the star is oxygen-rich,   PAH features produced by carbon-rich species at
 6.3 ${\mu}$m, 8.2 ${\mu}$m,  11.3  ${\mu}$m  and 12.7 ${\mu}$m are detected along with likely broad silicate emission near 20 ${\mu}$m. These results can be explained if there is  a passive, flared disk
orbiting HD 233517.   Our data support  the hypothesis that organic molecules in orbiting disks  may  be synthesized {\it in situ} as well
 as being incorporated from the interstellar medium.

\end{abstract}

\section{INTRODUCTION}
     
We  study  dusty circumstellar disks  in order to understand better the 
origin and  evolution of planets and
stars.  Analogous to young stellar objects, disks around evolved stars 
such as the Red Rectangle are important in the angular momentum budget 
of the system and also may be sites where planets form, as occurred 
around the pulsar
PSR 1215+12 (Wolszczan \& 
Frail 1992).  Here, we report data obtained with the 
Infrared Spectrograph (Houck et al. 2004) on the 
{\it Spitzer Space Telescope} (Werner et al. 2004) that
provide  supporting evidence for the hypothesis (Jura 2003) that 
the evolved red giant  HD 233517 (K2 III) possesses an orbiting disk.

About 10$^{5}$ red giants  were  detected with IRAS. While Asymptotic 
Giant Branch  (AGB) stars with luminosities larger than 10$^{3}$ L$_{\odot}$ 
usually have large enough mass-loss rates to produce substantial infrared 
excesses (see, for example, Sopka et al. 1985), the less powerful 
first-ascent red giants with luminosities
near 100 L$_{\odot}$ typically do not have measurable infrared excesses.  
Even when present, the typical fractional excess is 
 only between  10$^{-4}$ and 10$^{-3}$ of the star's luminosity 
 (Zuckerman et al. 1995), and at least in some cases actually may be produced 
 by interstellar cirrus rather than mass loss (Kim et al. 2001).
Only a handful of  class III red giants with larger  excesses  are known; 
a well-studied example  is  HD 233517 with a luminosity of 90 L$_{\odot}$ 
and a fractional excess of  ${\sim}$ 0.06 (Sylvester et al. 2001).  
Identified as anomalous in the IRAS
data base (Walker \& Wolstencroft 1988), initially, it was thought to be a 
main-sequence  star (Skinner et al. 1995).  Although its parallax is not 
measured, high-resolution optical spectroscopy shows that it is a  red 
giant (Fekel et al. 1996, Balachandran et al. 2000, Zuckerman 2001).  
Furthermore, HD 233517 is likely to be post-main-sequence rather than 
pre-main-sequence  because it is spatially isolated from any known region of star 
formation, does
not fall on any theoretical tracks in the H-R diagram for 
pre-main-sequence stars and shows an abundance of lithium of [Li]/[H] = 1.7 
${\times}$ 10$^{-8}$ about an order of magnitude greater than the standard 
interstellar value of 2 ${\times}$ 10$^{-9}$ (Anders \& Grevesse 1989, Howarth et al. 2002).  This large
abundance of lithium can be explained by models of surface mixing of
processed material in some models of post-main-sequence evolution 
(see, for example, Denissenkov \& Weiss 2000,  Denissenkov \& Herwig 2004), 
but not by current models for pre-main-sequence stars.     

 HD 233517 has  $v\,\sin\,i$ = 17.6 km s$^{-1}$ (Balachandran et al. 2000) which is substantially greater than the typical
rotational speed of a class III red giant of ${\leq}$5 km  s$^{-1}$ (de Medeiros \& Mayor 1995, Gray \& Pallavicini 1989, Schrijver \& Pols 1993), but smaller than the maximum possible of ${\sim}$35 km s$^{-1}$.  There are a few other first-ascent red
giants which also have marked infrared excesses, distinctively high lithium abundances and unusually rapid rotation (see, for example, Drake et al. 2002, Reddy \& Lambert 2005).   Any model to describe HD 233517 also may  pertain to these  stars.

Previously, Jura (2003) has argued that the  infrared excess around HD 233517 is unlikely to be produced by a recent outflow in a stellar wind.   For most  red giants that are currently losing mass,  F$_{\nu}$  typically varies as
${\nu}^{1.5}$ for  ${\lambda}$ $>$ 10 ${\mu}$m (see, for example, Sopka et al. 1985).  However, for HD 233517, the IRAS fluxes between  ${\lambda}$ = 13 ${\mu}$m and ${\lambda}$ = 60 ${\mu}$m vary  approximately as ${\nu}^{-5/3}$, a spectral energy 
distribution that can be naturally modeled with  a passive, flared, orbiting disk.  Also, the CO radio emission from the system is probably undetected (Jura \& Kahane 1999, Dent et al. 2005).  In contrast, the winds from  mass-losing red giants typically are easily detectable CO sources (see, for example, Olofsson et al. 1993).  
Here, we report additional evidence in support of the view that the infrared excess of HD 233517 is produced by orbiting dust.   

Our study of HD 233217 can help us understand other circumstellar disks.  
For example, there is controversy
about the origin of organic molecules in the early solar system.  The
usual view is that most of the carbonaceous material was incorporated as 
carbon-rich compounds from
the interstellar medium and then further processed (Ehrenfreund 
\& Charnley 2000,
Kerridge 1999).  However, a contrary view is that even in the oxygen-rich 
solar nebula,
Fischer-Tropsch catalysis on the surface of metal grains at $T$ $>$ 400 K
was an important route for the synthesis of carbon-rich molecules
(Kress \& Tielens 2001, Zolotov \& Shock 2001).  Alternatively, gas-phase 
synthesis of such molecules
 may be important (Morgan et al. 1991).  Since the disk around HD 233517 is free of interstellar contamination, it can serve as an indirect test of these models.

\section{OBSERVATIONS}
  	We observed HD 233517 on  2004 March 26  with the IRS; 
  	the data are labeled as AOR 3586048.  We operated the observatory in
  	 IRS Spectral Mapping mode where a 2 ${\times}$ 3 raster 
  	 (spatial x dispersion) centered on the star is performed.  
  	 The slit positions were separated by half of the slit width 
  	 in the dispersion direction, and by a third of the slit length 
  	 in the spatial direction.  We observed this object in both 
  	 orders of IRS Short-Low, SL,  (5.3-15 ${\mu}$m, 
  	 ${\lambda}$/${\Delta}{\lambda}$ ${\sim}$ 90).  
  	  Additionally, we observed this source in all 
  	  orders of IRS Short-High, SH, (10-20 ${\mu}$m, 
  	  ${\lambda}$/${\Delta}{\lambda}$ ${\sim}$  600) 
  	  and all orders of IRS Long-High, LH, 
  	  (20-40 ${\mu}$m, ${\lambda}$/${\Delta}{\lambda}$ ${\sim}$ 600).

	We reduced the spectra using the Spectral Modeling, Analysis, 
	and Reduction Tool (SMART; Higdon et al 2004).  We began with 
	the data product from the Spitzer Science Center 
	IRS calibration pipeline version S11.0.2 labeled `bcd', which 
	was already  processed with the stray light correction, and 
	the flatfield correction of the final basic calibrated data product.  
      The spectra for all modules came from reducing the data from 
      the center map positions.  For the SL data, we first subtracted sky 
      by subtracting the off-order of the same nod.  We then extracted 
      the spectra as a point-source for each nod position with a column 
      extraction about 4 pixels wide in the spatial direction.  The 
      spectra of each nod were then divided by the nodded spectrum of 
      our identically extracted photometric low-resolution standard star 
      ${\alpha}$ Lac, and then multiplied by the standard`s 
      template spectrum (Cohen et al 2003).  SH and LH data were 
      extracted using a full-slit extraction without any sky 
      subtraction due to a lack of off-order sky positions for 
      the high-resolution modules.  Each nod of the extracted 
      spectra was then divided by the corresponding nod of the 
      identically extracted high resolution photometric standard star 
      ${\xi}$ Dra, and then multiplied by the template spectrum.  We then merged 
      each nod of SL, SH, and LH to produce two complete spectra, one for each nod
            that were subsequently averaged.  We estimated the uncertainty 
            to be half the difference between the two nods.

      There are some small mismatches in flux between modules that are within 
      our absolute spectrophotometric accuracy of 10\%.  These mismatches were
       corrected by scaling the SH and LH modules to match the more reliably 
       calibrated SL.  Our reduced spectrum is shown in Figure 1.  Longward of 
       13 ${\mu}$m, the flux is a featureless continuum that rises to the red 
       with a kink near 20 ${\mu}$m.  Shortward of 13 ${\mu}$m, there are  PAH 
       features  at 6.26 ${\mu}$m, 8.24 ${\mu}$m, and 11.31 ${\mu}$m and  
       12.67 ${\mu}$m.  Figure 2 shows a more detailed view of the shorter 
       wavelength portion of the spectrum, and  Table 1 summarizes 
       the feature properties.
The feature wavelengths are well measured, but the energy carried in 
each
feature is uncertain because we cannot accurately measure the continuum level.
A particular uncertainty is that the broad feature which peaks at 
8.2 ${\mu}$m may actually be a blend of broad PAH emission (described as type C by Peeters et al. 2002)
which extends to
9.2 ${\mu}$m and silicate emission between 9.2 ${\mu}$m and
10.2 ${\mu}$m. 
The PAH features  carry about 1\% of 
of the total excess and thus re-radiate less than 0.1\% of the total stellar luminosity.   

\section{MODEL AND INTERPRETATION}

 In order to model the  continuum emission from the disk, we follow 
 Balachandran et al. (2000) and adopt a mass, $M_{*}$, of 1 M$_{\odot}$, a 
 radius, $R_{*}$, of 1.1 ${\times}$ 10$^{12}$ cm, an effective temperature, 
 $T_{*}$, of 4475 K, and a distance, $D_{*}$, of 620 pc.  
Following Jura (2003), we assume a passive, flared disk where a well-mixed 
fluid of  gas
and dust moves in circular orbits around the central star.  At every orbital 
radius, $R$, from the
star, the disk is assumed to have a single temperature which is controlled
by the balance between radiative heating of the disk from the star and re-radiation
of the disk in the infrared.  The gas and dust temperatures are assumed
equal to each other.  Perpendicular to the plane of the disk, vertical
hydrostatic equilibrium is established, and the  density distribution is described by a Gaussian with a half-thickness that is greater than the radius of the star. 

We first present a preliminary model to fit  the continuum called bb 
annulus \#2 in Figure 1,
 and we then describe an improved, final, model.
If $T_{disk}$ denotes the disk temperature, then:
\begin{equation}
T_{disk}\;=\;\left(\frac{1}{7}\right)^{2/7}\left(\frac{R_{*}}{R}\right)^{3/7}\,\left(\frac{2k_{B}T_{*}R_{*}}{GM_{*}{\mu}}\right)^{1/7}\;T_{*}
\end{equation}     
where $k_{B}$ is Boltzmann's constant, $G$ is the gravitational constant and
${\mu}$ is the mean molecular weight of the gas here taken as 3.9 ${\times}$ 10$^{-24}$ g.
The  flux at the Earth, $F_{\nu}$, is:
\begin{equation}
F_{\nu}\;=\;\frac{2\,{\pi}\,cos\,i}{D_{*}^{2}}\;{\int}_{R_{in}}^{R_{out}}\;B_{\nu}(T_{disk})\,R\,dR
\end{equation}
  To compute $F_{\nu}$, we need
to know the inner and outer boundaries of the disk, $R_{in}$ and $R_{out}$, and the inclination of the disk, $i$, which we take to equal 0$^{\circ}$.  Equations (1) and (2) yield:
\begin{equation}
F_{\nu}\;=\;\frac{28{\pi}}{3}\,\frac{cos\,i\,R_{*}^{2}}{D_{*}^{2}}\,\left(\frac{k_{B}T_{*}}{h{\nu}}\right)^{5/3}\frac{(k_{B}T_{*})^{3}}{(hc)^{2}}\left(\frac{2\,k_{B}T_{*}R_{*}}{49\,G\,M_{*}\,{\mu}}\right)^{2/3}\,{\int}_{x_{in}}^{x_{out}}\frac{x^{11/3}}{e^{x}\;-\;1}\,dx
\end{equation}
where $x$ = $h{\nu}/k_{B}T$.
From equation (3), we expect for an infinite disk that $F_{\nu}$ varies 
as ${\nu}^{-5/3}$ which matches the data reasonably well 
for ${\lambda}$ $>$ 20 ${\mu}$m.  However, it over-predicts the 
flux at ${\lambda}$ $<$ 20 ${\mu}$m and also does not match the kink in the
spectrum at this wavelength.  For our preliminary model to avoid producing 
too much shorter-wavelength radiation, we adopt an inner disk boundary of 
$R$ = 2.8 AU where $T$ = 230 K.  In order to account for the continuous
 rise in the emission to 35 ${\mu}$m, we require 
 that $T_{out}$ $<$ 65 K or that $R_{out}$ $>$  50 AU.   Ground-based observations yield an upper limit to the FWHM of the 18.2 ${\mu}$m  emission of  0{\farcs}41
(Fisher et al. 2003)  or a radius of  130 AU.  For our preliminary model, we adopt $R_{out}$ = 150 AU since the fluxes at ${\lambda}$ $<$ 35 ${\mu}$m are  insensitive to the exact value of $R_{out}$.  We 
see in Figure 1 that for ${\lambda}$ $>$ 13 ${\mu}$m, the preliminary model reproduces F$_{\nu}$ to within 20\%.

In the final model called bb annulus \#1 in Figure 1, we use 
a somewhat smaller dust torus by setting $R_{in}$ = 4 AU and $R_{out}$ = 63 AU and 
we add an ensemble of optically thin grains that we picture as suspended in 
the disk atmosphere  above and below the disk midplane.  This 
additional material includes 3 ${\times}$ 10$^{23}$ g of spherical 
pyroxene grains with radii of 0.1 ${\mu}$m at a temperature of 300 K, 
4.5 ${\times}$ 10$^{25}$ g of amorphous olivine with  radii of 
3 ${\mu}$m and a temperature of 130 K, and 4.5 ${\times}$ 10$^{25}$ g of 
amorphous pyroxene with radii of 3 ${\mu}$m and a temperature of 130 K.  
Finally, we scale the PAH emission reported by Sloan et al. (2005) for HD 
34282, a Herbig Ae/Be star; the template from this star provided the 
best match to the PAH features for HD 233517. As shown in Figure 1, the 
final model reproduces most of the the features near 10 ${\mu}$m and also 
better reproduces the kink in the continuum near 20 ${\mu}$m.

We can  estimate indirectly the amount of material in the system.  
In evaluating equation (3) to estimate the infrared flux from the disk 
and 
to produce the spectral fit seen in Figure 1, we assume a gas 
that is mostly H$_{2}$ and He.  Presuming that the initial outflow that 
produced the currently observed
disk is typical of mass-losing stars, we adopt
 a gas-to-dust ratio by mass of 100 and require enough dust to produce the 
 100 ${\mu}$m IRAS flux.  With these assumptions, the total gas mass in the 
 disk is  ${\sim}$0.01 M$_{\odot}$ and the  total disk angular momentum  is 
 ${\sim}$
3 ${\times}$ 10$^{51}$ g cm$^{2}$ s$^{-1}$, which is too much for  
a 
single star  of ${\sim}$1 M$_{\odot}$ (Jura 2003).  Initially, 
the system  must have been a binary.

 The PAH emission spectrum we report here for HD 233517 somewhat resembles  the
spectra for the Egg Nebula (or RAFGL 2688) and IRAS 13416-6243,  where the 
7.7 ${\mu}$m component is absent and the 8.2 ${\mu}$m component  is strong and 
asymmetric to the red (the rare 
class ``C" spectra, reported by Peeters et al. 2002).    Both the Nova V705 Cas (Evans et al. 2005) and
the pre-main-sequence star HD 135344 (Sloan et al. 2005) also exhibit
a pattern of PAH emission where the  7.7 ${\mu}$m component is relatively weak.

\section{DISCUSSION}

A striking result in  our data is the detection of PAH features 
in the circumstellar dust around  an oxygen-rich star. 
 If the  material was recently ejected from HD 233517, 
it would be mostly composed of silicates as found for 
almost all winds from oxygen-rich stars with a handful 
of possible exceptions (Sylvester et al. 1994, 1998, Speck et al. 2000). 
We interpret the presence of carbon-rich material in the circumstellar 
matter as evidence against a standard wind model. 

One unlikely hypothesis is that there is an unseen secondary star in the system that  was once  carbon-rich
and lost matter that  ended up in the disk around HD 233517.  Since mass-losing carbon stars on the AGB have luminosities larger than 10$^{3}$ L$_{\odot}$ (see, for example, Olofsson et al. 1993), this hypothetical mass-losing carbon star
must now be a white dwarf.  
The lack of radial velocity variation shows that there does not
appear to be any close stellar companion to HD 233517 (Balachandran et al. 2000). If there is a distant white dwarf at an orbital separation of, say, 50 AU, it is
conceivable that 1\% of its wind was captured by HD 233517 to produce a disk
of 0.01 M$_{\odot}$.  
  There are, however,  difficulties with the picture that HD 233517 has a distant
white dwarf companion.   First,  this model
does not offer an explanation for the very high lithium abundance in
HD 233517.  Second, this scenario does not require that
the mass-gaining star  be a red giant, but there are no known main-sequence 
stars with  infrared excesses similar to that of HD 233517.  Third, if there is a hot white
dwarf in the system near enough to transfer a substantial amount of matter
to the red giant, then the system might appear to be a symbiotic or at 
least display some unusual 
ultraviolet emission which is not seen.      

One possibility is that the PAH material around HD 233517 is the result of a set of 
Kuiper Belt Objects that are rapidly sublimating since the
star has become a red giant. A difficulty with this picture is that while comets 
contain a large amount of organic material, they apparently do not
possess many PAHs (see Ehrenfreund \& Charnley 2000) although there might be some 
as-yet unidentified chemical pathway to convert the organic material into PAHs. 
Also, we do not detect strong silicate emission characteristic of comets.

Following Jura (2003),  we hypothesize that while on the main sequence, HD 233517 was a short period binary.   When the more massive star of the pair became a red giant, there was an episode of rapid, unstable, mass transfer with the low-mass secondary spiraling into the primary.  When  the low-mass companion was engulfed, there was so much angular momentum in the system that mass was equatorially ejected into the immediate surroundings (see, for
example, Counselman 1973, Eggleton \& Kiseleva-Eggleton 2002, Hut 1980).  
The circumstellar matter subsequently expanded to its current size 
due to viscosity (see Pringle 1991).  Below,  we sketch a scenario by which this  disk might 
exhibit a carbon-rich chemistry.

At $T$ $<$
1000 K, the exact temperature depending upon the pressure, the thermodynamically preferred states of carbon and oxygen atoms
are in the molecules CH$_{4}$ and H$_{2}$O rather than CO 
(see, for example, Burrows \& Sharp 1999, Zolotov \& Shock 2001). At a sufficiently low temperature,  water and ice  droplets
 form    while CH$_{4}$ remains
in  the gas phase until the temperature drops significantly below 100 K (Boogert et al. 1998).   
The disk's approach to  this 
thermodynamically-predicted state in a sufficiently short time can be  
facilitated by Fischer-Tropsch (FT) catalysis on the 
surface of iron grains at phases when the material was warmer than 400 K,
 as has been  suggested to explain the presence of carbonaceous
material in meteorites (see, for example, Kress \& Tielens 2001, 
Zolotov \& Shock 2001). Alternatively, gas-phase chemistry may be 
important (Morgan et al. 1991).
   Although detailed models are 
required, we imagine a scenario somewhat similar to what   
 Willacy (2004) has computed  for the role of FT reactions in the 
 outflow
from the carbon-rich star IRC+10216 where CO 
and H$_{2}$ are converted to H$_{2}$O and hydrocarbons.  
She found that if solid iron grains are formed, then about 0.1\% of 
the CO is converted to water.   Since the 
density in the disk at, say 10 AU, around HD 233517 (${\sim}$10$^{12}$ 
cm$^{-3}$) is so much 
greater than in the warm region at 10 AU of the
wind from IRC+10216 (${\sim}$ 10$^{9}$ cm$^{-3}$), 
it is plausible
that  the FT reactions could effectively convert much of the carbon 
initially contained in CO into hydrocarbons around HD 233517. 
 Subsequently, perhaps analogous with  atmospheric chemistry  in
  Jupiter (Wong et al. 2003), these simple hydrocarbons could be 
  converted into PAHs.  Such a scenario may occur in other environments with 
  metal grains, 
  perhaps accounting for the apparent presence of PAHs in the circumstellar 
  shells around oxygen-rich
  red supergiants in $h$ and ${\chi}$ Persei (Sylvester et al. 1994).  
  The unusual character of the 
   PAH spectrum of HD 233517 may be a valuable but as yet unexplained clue 
   for understanding the 
  origin of this material in an oxygen-rich disk.
   
  \section{CONCLUSIONS}

The infrared spectrum of  HD 233517 shows emission from PAHs and has a 
 continuum that can be  reproduced approximately with a passive, flared, orbiting 
 disk.  The data can be explained if the material resides in a long-lived orbiting 
 disk which may have
been created when  HD 233517  engulfed a companion.  
The data lend support to the hypothesis that organic molecules in disks may  be 
synthesized {\it in situ} as well as being incorporated from the interstellar medium.

\newpage
\begin{figure}
\epsscale{1}
\plotone{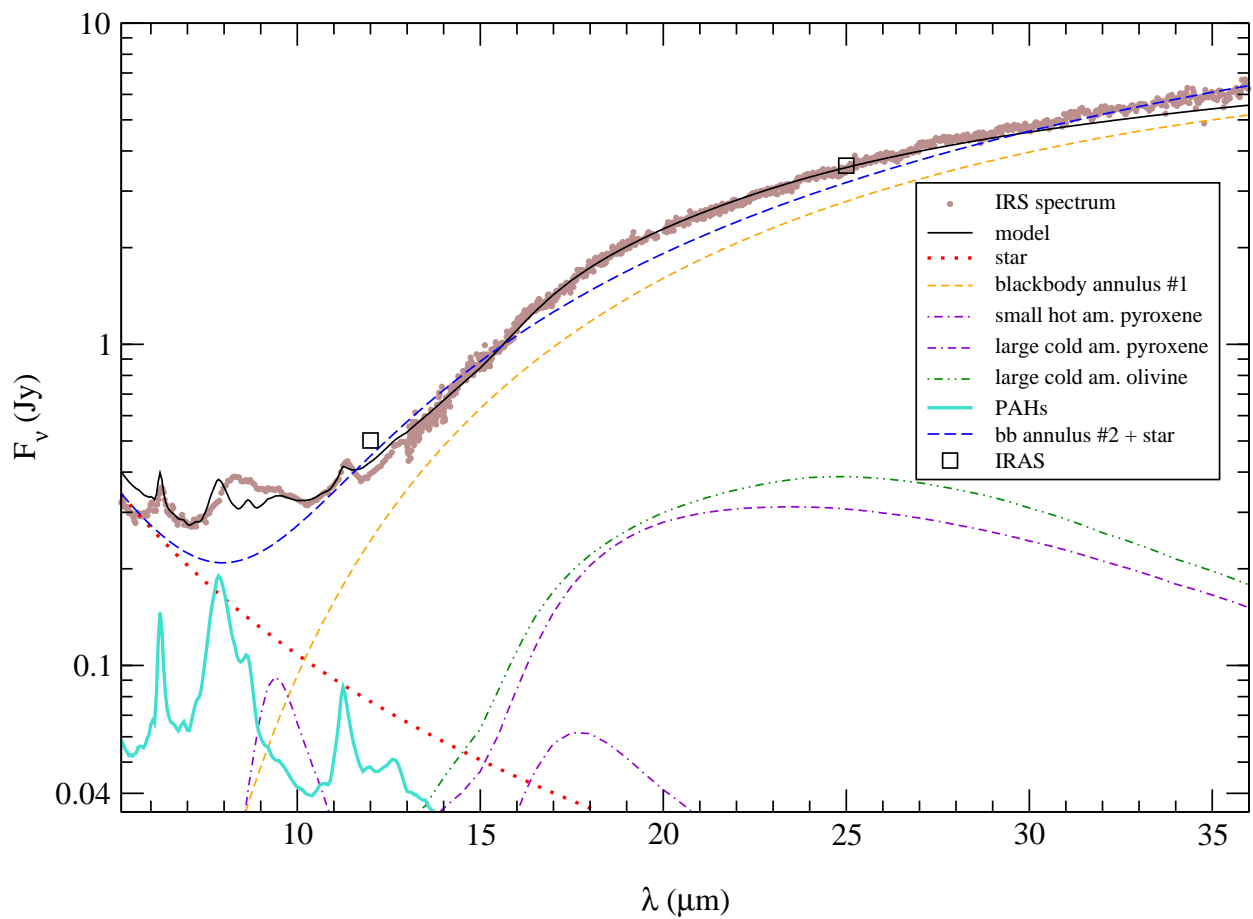}
\caption{The IRS spectrum of HD 233517 and the IRAS data points shown as open squares. The dashed line for bb annulus\# 2 + star denotes the preliminary model with only a flared passive opaque disk described in the text.
   The solid line shows the final model that includes both a somewhat smaller opaque disk (blackbody annulus \# 1), and a set of optically thin smaller particles including  pyroxene and olivine silicate grains with ``large" and ``small" defined in the text
  and the PAH emission scaled from HD 34282.  }
\end{figure}
\begin{figure}
\epsscale{1}
\plotone{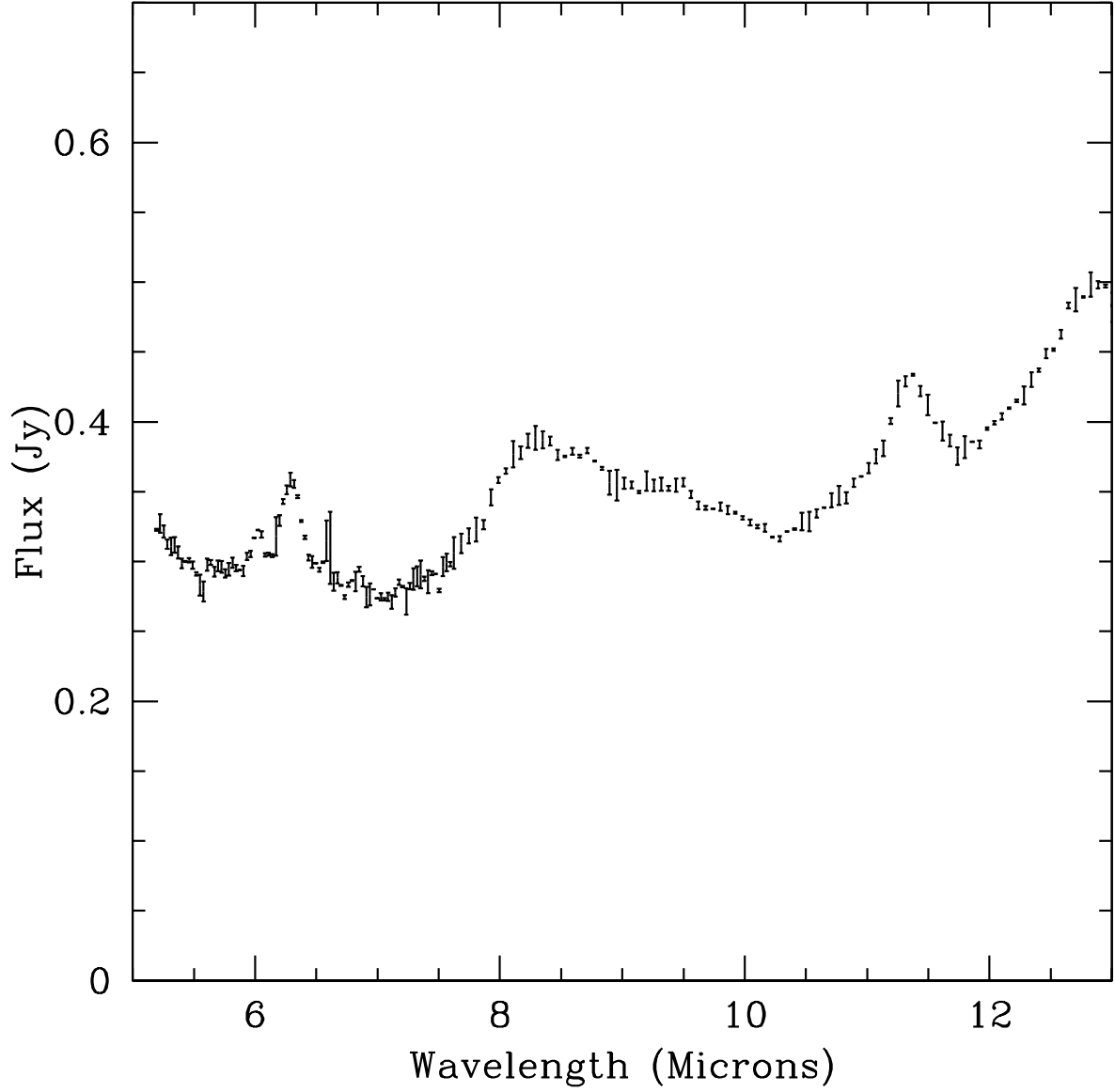}
\caption{An expanded version of the IRS spectrum of HD 233517 that shows the
emission features and their profiles together with the 1${\sigma}$ errors.  The apparent emission near 6.5 ${\mu}$m is not real.}
\end{figure}
\newpage
\begin{center}
Table 1 -- PAH Emission Features

\begin{tabular}{llll}
Wavelength  &  Line Flux  \\
\hline
\hline
(${\mu}$m) &   (10$^{-13}$ erg cm$^{-2}$ s$^{-1}$) \\
\hline
6.26 ${\pm}$ 0.019 & 19 ${\pm}$ 2     \\
8.24 ${\pm}$ 0.037  & 24 ${\pm}$ 2  \\
11.31 ${\pm}$ 0.022   & 7 ${\pm}$ 1  \\ 
12.67 ${\pm}$ 0.031 & 4 ${\pm}$ 1 \\  
\hline
\end{tabular}
\end{center}

\begin{thebibliography}{}
\bibitem{anders89} Anders, E. \& Grevesse, N. 1989, Geochim. Cosmochim. Acta, 53, 197
\bibitem{bala00} Balachandran, S. C., Fekel, F. C., Henry, G. W. \& Uitenbroek, H., 2000, \apj, 542, 978
\bibitem{boogert98} Boogert, A. C. A., Helmich, F. P., van Dishoeck, E. F., Schutte, W. A., Tielens, A. G. G. M. \& Whittet, D. C. B. 1998, \aap, 336, 352
\bibitem{burrows99} Burrows, A. \& Sharp, C. M. 1999, \apj, 512, 843
\bibitem{cohen03} Cohen, M., Megeath, S. T., Hammersley, P. L., Martin-Luis, F. \& Stauffer, J. 2003, \aj, 125, 2645
\bibitem{counselman73} Counselman, C. C. 1973, \apj, 180, 307
\bibitem{demedeiros95} de Medeiros, J. R. \& Mayor, M. 1995, \aap, 302, 745
\bibitem{denissenkov00} Denissenkov, P. A. \& Weiss, A. 2000, \aap, 358, L49
\bibitem{denissenkov04} Denissenkov, P. A. \& Herwig, F. 2004, \apj, 612, 1081
\bibitem{dent05} Dent, W. R. F., Greaves, J. S. \& Coulson, I. M. 2005, \mnras, 359, 663
\bibitem{drake02} Drake, N. A., de la Reza, R., da Silva, L. \& Lambert, D. L.
2002, \aj, 123, 2703
\bibitem{eggleton02} Eggleton, P. P. \& Kiseleva-Eggleton, L. 2002, \apj, 575, 461
\bibitem{ehrenfreund00} Ehrenfreund, P. \& Charnley, S. B. 2000, \araa, 38, 427
\bibitem{evans05} Evans, A., Tyne, V. H., Smith, O., Geballe, T. R., Rawlings, J. M. C. \& Eyres, S. P. S. 2005, \mnras, 360, 1483
\bibitem{fekel96} Fekel, F. C., Webb, R. A., White, R. J. \& Zuckerman, B. 1996, \apj, 462, L95
\bibitem{fisher03} Fisher, R. S., Telesco, C. M., Pina, R. K. \& Knacke, R. F. 2003, \apj, 586, 91
\bibitem{gray89} Gray, D. F. \& Pallavicini, R.  1989, \pasp, 101, 695
\bibitem{higdon04} Higdon, S. J. U. et al. 2004, \pasp, 116, 975
\bibitem{houck2004} Houck, J. 2004 et al., \apjs, 154, 18
\bibitem{howarth02} Howarth, I. D., Price, R. J., Crawford, I. A. \& Hawkins, I. 2002, \mnras, 325, 267
\bibitem{hut80} Hut, P. 1980, \aap, 92, 167
\bibitem{jura03} Jura, M. 2003, \apj, 582, 1032
\bibitem{jura99} Jura, M. \& Kahane, C. 1999, \apj, 521, 302
\bibitem{kerridge99} Kerridge, J. F. 1999, Space Sci Rev, 90, 275
\bibitem{kim2001} Kim, S. S., Zuckerman, B. \& Silverstone, M. 2001, \apj, 550, 1000
\bibitem{kress} Kress, M. E., \& Tielens, A. G. G. M. 2001, Met. \& Plan. Sci., 36, 75
\bibitem{morgan91} Morgan, W. A., Feigelson, E. D., Wang, H. \& Frenklach, M. 1991, Science, 252, 109
\bibitem{olofsson93} Olofsson, H., Eriksson, K., Gustafsson, B. \& Carlstrom, U.
1993, \apjs, 87, 267
\bibitem{peeters02} Peeters, E., Hony, S., Van Kerckhoven, C., Tielens, A. G. G. M., Allamandola, L. J., Hudgins, D. M. \& Bauschlicher, C. W. 2002, \aap, 390, 1089
\bibitem{pringle91} Pringle, J. E. 1991, \mnras, 248, 754
\bibitem{reddy05} Reddy, B. E. \& Lambert, D. L. 2005, \aj, 129, 2813
\bibitem{schrijver93} Schrijver, C. J. \& Pols, O. R. 1993, \aap, 278, 51
\bibitem{skinner94} Skinner, C. J. 1994, \mnras, 271, 300
\bibitem{skinner95} Skinner, C. J., Sylvester, R. J., Graham, J. R., Barlow, M. J., Meixner, M., Keto, E., Arens, J. F. \& Jernigan, J. G. 1995, \apj, 444, 861
\bibitem{sloan05} Sloan, G. C. et al. 2005, \apj, 632, 956
\bibitem{sopka85} Sopka, R. J., Hildebrand, R., Jaffe, D. T., Gatley, I., Roellig, T., Werner, M., Jura, M. \& Zuckerman, B. 1985, \apj, 294, 242
\bibitem{speck00} Speck, A., K., Barlow, M. J., Sylvester, R. J. \& Hofmeister, A. M. 2000, \aaps, 146, 437
\bibitem{sylvester94} Sylvester, R. J., Skinner, C. J. \& Barlow, M. J. 1994, \mnras, 266, 640
\bibitem{sylvester98} Sylvester, R. J., Skinner, C. J. \& Barlow, M. J. 1998, \mnras, 301, 1083
\bibitem{sylvester01} Sylvester, R. J., Dunkin, S. K. \& Barlow, M. J. 2001, \mnras, 327, 133
\bibitem{Walker1988} Walker, H. J. \& Wolstencroft, R. D. 1988, PASP, 100, 1509
\bibitem{werner2004} Werner, M. W. et al. 2004, \apjs, 154, 1
\bibitem{willacy2004} Willacy, K. 2004, \apj, 600, L87 
\bibitem{wolszczan92} Wolszczan, A. \& Frail, D. A. 1992, Nature, 355, 145
\bibitem{wong03} Wong, A.-S., Yung, Y. L., \& Friedson, A. J. 2003, Geophys. Res. Lett., 30, 30.
\bibitem{zolotov2001} Zolotov, M. Y. \& Shock, E. L. 2001, Icarus, 150, 323
\bibitem{zuckerman2001} Zuckerman, B. 2001, \araa, 39, 549
\bibitem{zuckerman1995} Zuckerman, B., Kim, S.-S. \&Liu, T. 1995, \apj, 446, L79
\end{thebibliography}
\end{document}